\documentclass[fleqn,10pt]{wlscirep}

\usepackage[utf8]{inputenc}
\usepackage[T1]{fontenc}
\title{Structured transmittance illumination coherence holography}

\author[1,2]{Aditya Chandra Mandal}
\author[1]{Tushar Sarkar}
\author[3]{Zeev Zalevsky}
\author[1,*]{Rakesh Kumar Singh}
\affil[1]{Laboratory of Information Photonics and Optical Metrology, Department of Physics, Indian Institute of Technology (Banaras Hindu University), Varanasi, 221005, Uttar Pradesh, India}
\affil[2]{Department of Mining Engineering, Indian Institute of Technology (Banaras Hindu University), Varanasi, 221005, Uttar Pradesh, India}
\affil[3]{Bar-Ilan University, Faculty of Engineering and Nano Technology Center, Ramat-Gan, Israel}

\affil[*]{krakeshsingh.phy@iitbhu.ac.in}

\usepackage{subfig}
\usepackage{float}
\usepackage{graphicx}
\usepackage[autostyle]{csquotes}

\begin{abstract}

The coherence holography offers an unconventional way to reconstruct the hologram where an incoherent light
illumination is used for reconstruction purposes, and object encoded into the hologram is reconstructed as the distribution
of the complex coherence function. Measurement of the coherence function usually requires an interferometric setup and array detectors.
This paper presents an entirely new idea of reconstruction of the complex coherence function in the coherence
holography without an interferometric setup. This is realized by structured pattern projections on the incoherent source
structure and implementing measurement of the cross-covariance of the intensities by a single-pixel detector. This technique, named structured transmittance
illumination coherence holography (STICH), helps to reconstruct the complex coherence from the intensity
measurement in a single-pixel detector without an interferometric setup and also keeps advantages of the intensity
correlations. A simple experimental setup is presented as a first step to realize the technique, and results based on the
computer modeling of the experimental setup are presented to show validation of the idea.
\end{abstract}
\begin{document}

\flushbottom
\maketitle
%
%
\thispagestyle{empty}

\section*{Introduction}

Since its inception more than seventy years ago, holography offers a powerful tool for imaging and light synthesis \cite{gabor1948new,leith1962reconstructed,yamaguchi2006phase,hendry2006digital}. The availability of high-quality array detectors and reconstruction algorithms has further revolutionized the holography, and optical reconstruction in the holography is replaced by the digital means, and technique is called digital holography (DH).
The DH keeps inherent advantages of the holography concerning the complex amplitude distribution and additionally offers a simplified reconstruction of the hologram by a numerical means\cite{hendry2006digital}. Conventional DH records and reconstructs the complex wavefield by an optical field distribution itself. The phase information of the wavefield is an important physical parameter of the light. Among various methods to recover the phase information, the DH is a well-established technique for quantitative phase imaging (QPI). Various QPI 
techniques have been developed, and significant among them are in-line, off-axis, and phase-shifting holography. In recent years, attempts have been made to improve the transverse spatial resolution in the DH by random \cite{park2009speckle,vinu2019speckle,choi2014dynamic} and structured light illumination\cite{gao2013structured,ma2013resolution,mico2019resolution,gao2021recent}. Requirements to retrieve information from the self-luminous or incoherent object have also inspired new trends in the holography \cite{rosen2007digital,tahara2017single,bouchal2011point,rosen2007fluorescence,quan2017single,watanabe2015recording,rosen2009review}.

In a significant development, Takeda and co-workers have developed an unconventional holography called coherence holography (CH), where the information of the complex field is reconstructed as a distribution of the spatial coherence\cite{takeda2005coherence}. Here, the hologram is reconstructed by an incoherent light illumination rather than by a coherent light. The CH has opened new research directions on recording and shaping the spatial coherence for applications such as spatial coherence tomography, profilometry, imaging and coherence current\cite{duan2006dispersion, takeda2014spatial,naik2012coherence,soni2016polarization,wang2006coherence}. Principle of the CH is derived from the van Cittert-Zernike theorem, which connects the incoherent source structure with a far-field spatial coherence of the light. The main task in the CH is to design an appropriate
interferometer for the measurement of spatial coherence. These interferometers mainly employ second-order correlation or fourth-order correlations measurement by the array detectors. In contrast to the interferometers based on the second-order correlations, the fourth-order correlation, i.e., intensity interferometers are highly stable for the coherence measurement\cite{goodman2015statistical}. Recently, Naik et al.\cite{naik2011photon} made use of the fourth-order correlation in the hologram, and technique is called a photon correlation holography (PCH). Basic principle of the PCH is derived from a connection between the cross-covariance of the
intensities with the modulus square of the Fourier spectrum of the incoherent structure. However, phase information of
the spectrum is lost in the PCH in contrast to the CH and DH techniques. Recovery of complex field parameters in the CH
and DH is possible by employing an interferometer setup which makes the system bulky and prone to external
disturbances and instabilities\cite{leith1962reconstructed,soni2016polarization,takeda2014spatial}.
Recently, holographic methods based on interference of the coherence waves have been proposed to overcome the phase loss issue in the PCH experiments\cite{kumar2014recovery,singh2017lensless,kim2019imaging,somkuwar2017holographic,chen2020phase,chen2022increasing}. However, the  interferometric systems bring bulkiness in the experimental implementation and also require  flexible control over the reference field to get the interference fringes in the cross-covariance function at the array detectors plane.

On the other hand, significant attempts have been made to develop computational imaging techniques such as single-pixel imaging with random and structured field illumination over the past few years\cite{zhang2015single,martinez2017single,horisaki2017single,edgar2019principles,shin2018reference,hu2019single,ota2018complex,singh2017hybrid,gibson2020single}. In contrast to using
conventional cameras and two-dimensional array detectors, single-pixel techniques make use of projection of light
patterns onto a sample while a single-pixel detector measures the light intensity collected for each pattern. Therefore,
stage of spatial sampling is moved from the camera to the programmable diffraction element where the structured
patterns are loaded. Single-pixel imaging techniques have brought advantages such as use of a non-visible wavelength or precise
time resolution, which can be costly and practically challenging to realize as a pixilated imaging device. Recently, a combination of optical and computation channels has been developed for the reconstruction of the three-dimensional (3D) amplitude object from
a single-pixel detector, and technique is called hybrid correlation holography (HCH)\cite{singh2017hybrid}. This technique makes use of
cross-covariance of the intensity and is derived from the connection between complex coherence function and
intensity correlation for Gaussian random field. A new scheme based on the recovery of the complex-valued object in a
modified HCH scheme with an interferometric setup has been developed\cite{chen2020complex}.

In this paper, we present a new technique for the reconstruction of the complex field within the framework of the PCH
and present a new theoretical basis for the reconstruction in correlation holography. This approach equips
correlation imaging with a complete wavefront reconstruction without an interferometric setup but keeping the advantage
of the intensity correlation. For this purpose, a structured light illumination is projected on the incoherent structure, and
a far-field spectrum is measured by a single-pixel detector. A complex Fourier spectrum from the
intensities is successfully obtained from the four-step phase shifting in the structured illumination. Although
the Fourier spectrum measurement in the CH is based on the Hanbury Brown-Twiss (HBT) approach with a single-pixel
detector but active illumination strategy in the proposed technique helps to overcome the phase loss problem of the
typical HBT approach. Applying two-dimensional (2D) Fourier transform (IFT) to the
obtained spectrum yields the desired DH. The phase-shifting illumination approach also brings the elimination of noise that is statistically the same. A detailed theoretical foundation and implementation of the proposed technique in
comparison to the CH and PCH are discussed below.
\section*{Basic Principle}
Basic principles of the CH and PCH have been discussed in detail in Ref.\cite{takeda2005coherence,naik2011photon}. However, for the sake of continuity and to connect with basic principle of the proposed technique, we briefly describe the CH and PCH. Fig. 1(a) represents a coherent recording of the complex field of an object in the Fourier hologram  $ H(r) $. The hologram $ H(r) $  is read out with the incoherent light, as shown in Fig. 1(b). To describe the reconstruction process of the hologram in Fig. 1(b), consider the complex field of light immediately behind the hologram as
\begin{equation}\label{1}
 E(r)=H(r) \exp [i \phi(r)]   
\end{equation}
where $i$ denotes the imaginary unit and $H(r)=|H(r)| \exp [i \delta(r)]$ with $|H(r)|$ and $\delta(r)$ being the amplitude transmittance of the hologram and deterministic phase of the readout light, respectively. The spatial vector at the source is $r\equiv(x,y)$. The random phase inserted in the light path to destroy spatial coherence by the rotating ground glass (RGG) is represented by $\phi(r)$ at a fixed time $t$. A lens in Fig. 1(b) with focal distance $f$ is used to Fourier transform the randomly scatted light field from the source, and the complex field on the observation plane becomes 
\begin{equation}\label{2}
    E(k)=\int E(r) \exp  \left[-i\enspace{k \cdot r}\right] d r=\int H(r) \exp [i \phi(r)]\exp \left[-i\enspace k \cdot r\right] d r
\end{equation}
where $k \equiv (k_{x},k_{y})$ is spatial frequency coordinate at the observation point. Two-point correlation of the random field is characterized as
\begin{equation}\label{3}
\begin{split}
       W\left(k_{1}, k_{2}\right) & =\left\langle E^{*}\left(k_{1}\right) E\left(k_{2}\right)\right\rangle \\
    & =\iint H^{*}\left(k_{1}\right) H\left(k_{2}\right)\langle\exp \left[i\left(\phi\left(k_{2}\right)-\phi\left(k_{1}\right)\right)\right]\rangle\exp \left[-i \left(k_{2} \cdot {r_{2}}-k_{1} \cdot r_{1}\right)\right] d r_{2} d {r_{1}}
\end{split}
\end{equation}
here $\left\langle.\right\rangle$ represents the ensemble average which will be replaced  by the temporal average in the experiment. The rotating ground glass is considered to produce an incoherent source, i.e. $\langle\exp \left[i\left(\phi\left(r_{2}\right)-\phi\left(r_{1}\right)\right)\right]\rangle \equiv \delta\left(r_{2}-r_{1}\right)$, where $k_{2}=k$ and $k_{1}=0$. Therefore, Eq. \ref{3} transforms into the van Cittert Zernike theorem as
\begin{equation}\label{4}
    F(k)=\int I(r) \exp \left[-i\enspace k \cdot r\right] d r
\end{equation}
where $I(r)=H^{*}(r)H(r)$  is the source hologram placed at the RGG plane and $F(k)$ represents the Fourier spectrum of the incoherent source at the far-field. The basic principle of the CH is described by Eq. \ref{4} and therefore provides reconstruction of the object as the distribution of the complex coherence function. 
The random field intensity at the observation plane, at a fixed time $t$ corresponding to one rotation state of the RGG, is represented as 
\begin{equation}\label{5}
    I(k)=|E(k)|^{2}
\end{equation}

\begin{figure}[]
  \centering
  \begin{tabular}[b]{c}
    \includegraphics[width=.35\linewidth]{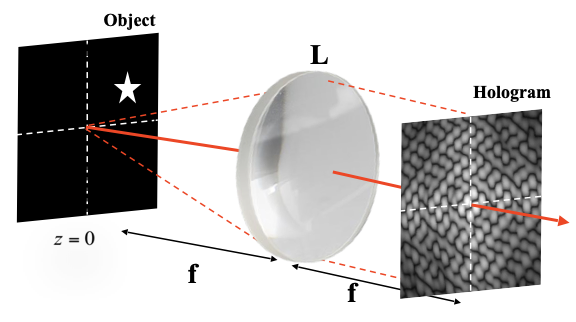} \\
    \small (a)
  \end{tabular} \enspace
  \begin{tabular}[b]{c}
    \includegraphics[width=.45\linewidth]{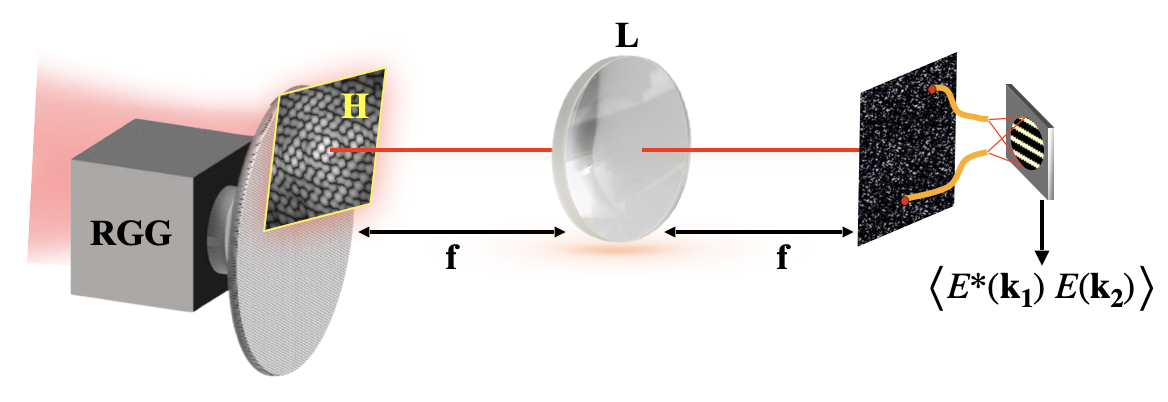} \\
    \small (b)
  \end{tabular} \enspace
   \begin{tabular}[b]{c}
    \includegraphics[width=.45\linewidth]{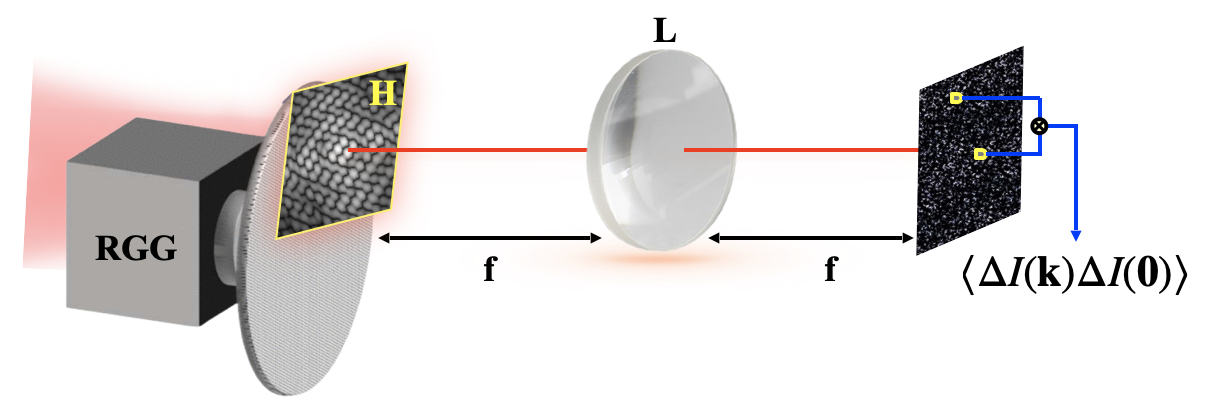} \\
    \small (c)
  \end{tabular}
  \begin{tabular}[b]{c}
    \includegraphics[width=.45\linewidth]{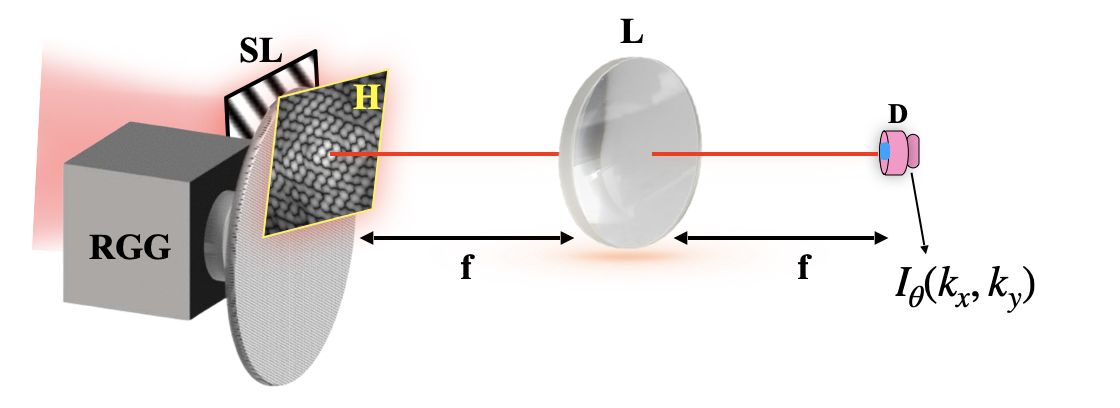} \\
    \small (d)
  \end{tabular}
  \caption{(a) Represents coherent recording of complex field of an object located at $Z=0$  , $L$ is a lens with focal distance $f$ (b) Represents the reconstruction process of the hologram with incoherent light, where $L$ is a lens with focal distance $f$, RGG is rotating ground glass, $H$ is hologram ; (c) Represents the PCH setup: intensity correlation holography scheme; (d) Represents the STICH setup, where $SL$ is structured light pattern, $H$ is hologram, $D$ is single-pixel detector}
\end{figure}
The random intensity pattern $I(k)$ is having no direct resemblance to reconstruction of the object. The cross-covariance of the intensities of the Gaussian random field is given as
\begin{equation}\label{6}
   \langle\Delta I(k) \Delta I(0)\rangle=|F(k)|^{2}
\end{equation}
where $ \Delta I(k)=I(k)-\langle\Delta I(k)\rangle$ is the fluctuation of the intensities with respect to its average mean value. Eq. \ref{6} highlights the basic principle of the PCH and is sketched in Fig. 1(c), wherein the phase part of the coherence function is lost.
To circumvent the above-mentioned issue on recovery of complex phase with the only measurement of the cross-covariance of the intensities, we present a new technique called STICH. Basic principle of the STICH is represented in Fig. 1(d) and described as follows.
A two dimensional (2-D) structured illumination with its spatial frequency $(k_{x},k_{y})$ and initial phase $\theta$ is projected on the RGG. This structured illumination is a sinusoidal pattern and is represented as
\begin{equation}\label{7}
    P_{\theta}\left(x, y ; k_{x}, k_{y}\right)=a+b \cdot \cos \left( k_{x} x+ k_{y} y+\theta\right)
\end{equation}
where $a$ is an un-modulated term of the illumination pattern, and $b$ represents the contrast. The light coming out of the structured transparency propagates through the RGG, which is used to mimic an incoherent light source. 
A hologram $H$ is placed next to the RGG, as shown in Fig. 1(d). Therefore, the instantaneous complex field immediately after the $H$ is expressed as:
\begin{equation}\label{8}
 E_{\theta}\left(x, y ; k_{x}, k_{y}\right)=H(x, y) \exp (i \phi(x, y))\left(a+b \cdot \cos \left( k_{x} x+ k_{y} y+\theta\right)\right)
\end{equation}
The instantaneous complex field at the single-pixel detector is represented as
\begin{equation}\label{9}
E_{\theta}\left(k_{x}, k_{y}\right)=E_{n}+w \iint_{\Omega} E_{\theta}\left(x, y ; k_{x}, k_{y}\right) d x d y
\end{equation}
where $\Omega$ represents the illuminated area, $w$ is a scale factor whose value depends on the size and the location of the detector, $E_{n}$ represents the response of background illumination.
The instantaneous random intensity at the single-pixel detector is given as\begin{equation}\label{10}
I_{\theta}\left(k_{x}, k_{y}\right)=|E_{\theta}\left(k_{x}, k_{y}\right)|^{2}
\end{equation}
The random intensity variation from its mean intensity is calculated as
\begin{equation}\label{11}
\Delta I_{\theta}\left(k_{x}, k_{y}\right)=I_{\theta}\left(k_{x}, k_{y}\right)-\langle I_{\theta}\left(k_{x}, k_{y}\right)\rangle
\end{equation}
 where the angular bracket $\langle . \rangle $ denotes the ensemble average and  $\langle I_{\theta}\left(k_{x}, k_{y}\right)\rangle$ is mean intensity. The cross-covariance of the intensities is 
 \begin{equation}\label{12}
 D_{\theta}\left(k_{x}, k_{y}\right)=\langle \Delta I_{\theta}\left(k_{x}, k_{y}\right)\Delta I_{\theta}\left(k_{x}, k_{y}\right)\rangle
\end{equation}
The 4-step phase-shifting approach allows each complex Fourier coefficient $F(k_{x},k_{y})$ to be obtained by every four responses corresponding to the illumination patterns, i.e. $ P_{0}$, $P_{\pi/2}$, $P_{\pi/2}$ , $P_{3\pi/2}$. 
The response $D_{\theta}(\theta=0,\pi/2,\pi,3\pi/2)$ are used to obtain the Fourier spectrum of $H$, i.e. $F(k_{x},k_{y})$ as
\begin{equation}\label{13}
    \begin{array}{l}
{\left[D_{0}\left(k_{x}, k_{y}\right)-D_{\pi}\left(k_{x}, k_{y}\right)\right]+i \cdot\left[D_{\pi / 2}\left(k_{x}, k_{y}\right)-D_{3 \pi / 2}\left(k_{x}, k_{y}\right)\right]} \\
=2 b w \iint_{\Omega} H(x, y) \cdot \exp \left[-i \left(k_{x} x+k_{y} y\right)\right] d x d y
\end{array}
\end{equation}
The Fourier coefficient is expressed as
\begin{equation}\label{14}
    \begin{array}{l}
F\left(k_{x}, k_{y}\right)=\iint_{\Omega} H(x, y) \cdot \exp \left[-i \left(k_{x} x+k_{y} y\right)\right] d x d y \\
=\frac{1}{2 b w}\left[D_{0}\left(k_{x}, k_{y}\right)-D_{\pi}\left(k_{x}, k_{y}\right)\right]+i \cdot\left[D_{\pi / 2}\left(k_{x}, k_{y}\right)-D_{3 \pi / 2}\left(k_{x}, k_{y}\right)\right]
\end{array}
\end{equation}
By computing Fourier coefficients $F$ (i.e., the Fourier spectrum) using Eq. \ref{14} for a complete set of $(k_{x},k_{y})$, the desired complex field  distribution is reconstructed.
A 4-step phase-shifting sinusoid illumination plays an essential role in the proposed technique. Eq. \ref{13} can not only assemble the Fourier spectrum of the desired hologram image but also eliminate undesired  direct current (DC) terms. 
\section*{Experimental design and algorithm}
A possible experimental design for the proposed technique is shown in Fig. 2. A monochromatic collimated laser light is folded by a beam splitter (BS) and incident on a spatial light modulator (SLM). The SLM in Fig. 2 is considered to be a reflective type and loaded with the $h$ number of sinusoidal gratings in a sequence.
\begin{figure}[h]
\centering\includegraphics[width=13cm]{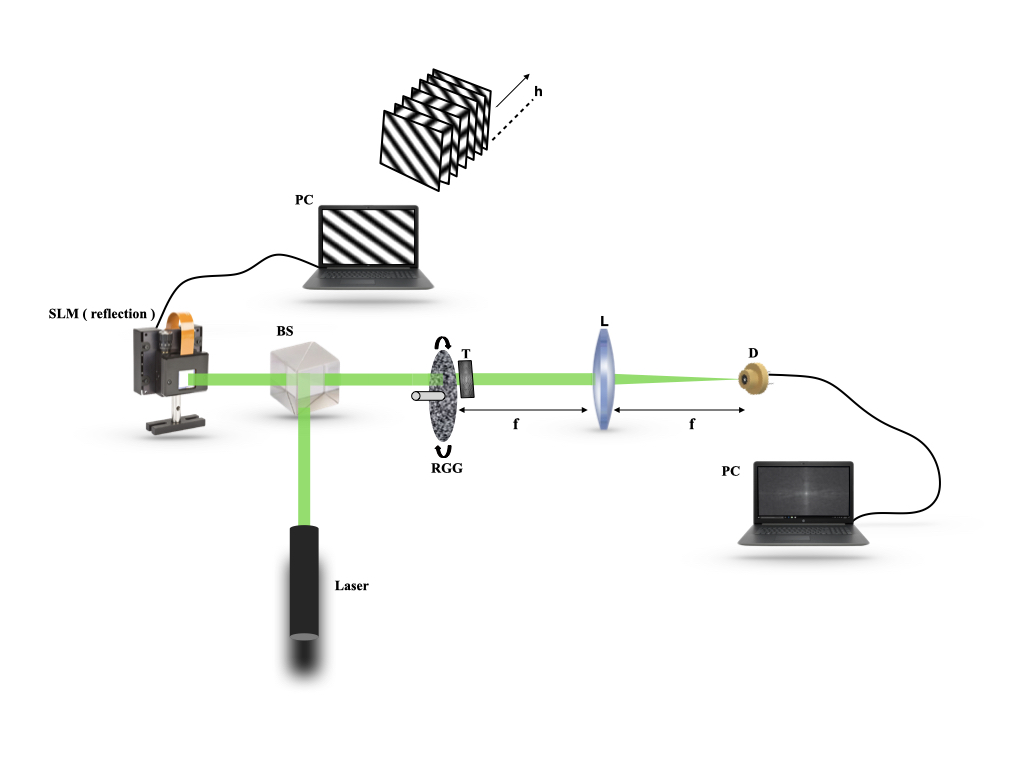}
\caption{Experimental configuration: BS- beam splitter, SLM- spatial light modulator, RGG- rotating ground glass, T-   transparency, L- lens with focal distance $f$, D- single-pixel detector, PC- personal computer.}
\end{figure}
The sinusoidal pattern displayed to the SLM is inserted into the incident beam, and subsequently, this structured light transmits through the BS and illuminates the RGG. The RGG introduces randomness in the incident structured light. As shown in Fig. 1(a), a computer-generated hologram of an off-axis object is used as transparency and placed adjacent to the RGG. The structured pattern embedded in the stochastic field due to the RGG illuminates a hologram $H(x,y)$. Scattering of the light through the RGG generates a stochastic field with the Gaussian statistics. The scattered light further propagates and is Fourier transformed by a lens $L$ at the single-pixel detector plane $D$.  Corresponding to the spatial frequency and initial phase of the loaded structured pattern, an instantaneous random field at the single-pixel detector is represented by $ E_{\theta}\left(k_{x}, k_{y}\right)$. The instantaneous signal at the detector is represented as  $|E_{\theta}\left(k_{x}, k_{y}\right)|^{2}$. After the single-pixel measurement for a particular random phase mask, we stored the value in our personal computer (PC) for post-processing. Due to the Gaussian statistics, the cross-covariance of the intensities is estimated at the single-pixel corresponding to different sets of random phase masks introduced by the RGG. The cross-covariance of the intensities at the single-pixel detector for a given frequency pair $(k_{x}, k_{y})$ and initial phase $\theta$ is represented as $ D_{\theta}\left(k_{x}, k_{y}\right)$. For a complete set of Fourier coefficients, we illuminate the hologram $H(x,y)$ by the structured patterns with full sets of spatial frequency $(k_{x}, k_{y})$. Each complex Fourier coefficient corresponding to that spatial frequency is extracted by using a 4-step phase-shifting approach. The number of Fourier coefficients in the Fourier domain is the same as the number of pixels in the spatial domain.

The algorithm proposed in this paper is an iterative heuristic that aims to reconstruct complex object encoded into the hologram as the distribution of the complex coherence function. In contrast to the previously reported CH, here we report the use of the structured illumination at the RGG plane. This strategy makes reconstruction procedure completely different from previously developed reconstruction approach and also equips us to extract the complex field even from a single-pixel detector. This is a unique feature of our proposed technique that helps to recover the complex coherence without interferometry. The algorithm in our work is implemented using MATLAB and simulated on a personal computer.
Fig. 3 shows the steps of the algorithm, which are:
\begin{enumerate}
    \item A hologram of size $N \times N$ pixels is taken as transparency.
    \item Construction of sinusoidal patterns and random phase masks :  
    \begin{enumerate}
        \item Total $M$ number of different random phase masks of same size of hologram are generated.
        \item The 2-D sinusoidal patterns of size  $N \times N$  pixels with initial phase  ${\theta}=(0,\pi/2,\pi,3\pi/2)$ are constructed by considering discretized spatial frequency space $-k<k_{x}, k_{y}<k$.
    \end{enumerate}
    \item Iterative steps:
    \begin{enumerate}
        \item First, a 2-D sinusoidal pattern for that particular frequency pair $(k_{x}, k_{y})$ is taken. A single-pixel detector is used to sense the random light field ( matrix multiplication of hologram and sinusoidal pattern and random phase mask ), and corresponding random intensity is represented using Eq. \ref{10}.
        \item In such a way, other intensity patterns at single-pixel detector corresponding to different sets of random phase masks introduced by the RGG are obtained. Random intensity variations from its mean intensity are calculated using Eq. \ref{11}. Cross-covariance of the intensity for the taken spatial frequency pair $(k_{x}, k_{y})$ is calculated using Eq. \ref{12}.
        \item Calculation of each complex valued Fourier coefficient $F\left(k_{x}, k_{y}\right)$ corresponding to the spatial frequency pair is simulated from four responses $D_{\theta}(\theta=0,\pi/2,\pi,3\pi/2)$ ( see Eq. \ref{14}). 
    \end{enumerate}
For the complete set of desired $R\times R$ size Fourier coefficients, we need to iterate the above steps over the discretized spatial frequency space unless the last iteration is obtained.
Total $4R^{2}(=4 \times R \times R)$ number of sinusoidal patterns need to be projected, including the four-step phase-shifting by an SLM shown in Fig. 2, on which the patterns are controlled by a personal computer (PC) directly. A Fourier coefficients $F\left(k_{x}, k_{y}\right)$ is obtained with 4 measurements. So, basically, for fully sampling $R \times R$ size Fourier coefficients consumes $4 \cdot R^{2} \cdot M(=4 \times R \times R \times M)$ measurements of single-pixel detector to reconstruct the complex object.
    
\end{enumerate}
\begin{figure}[H]
\centering\includegraphics[width=13cm]{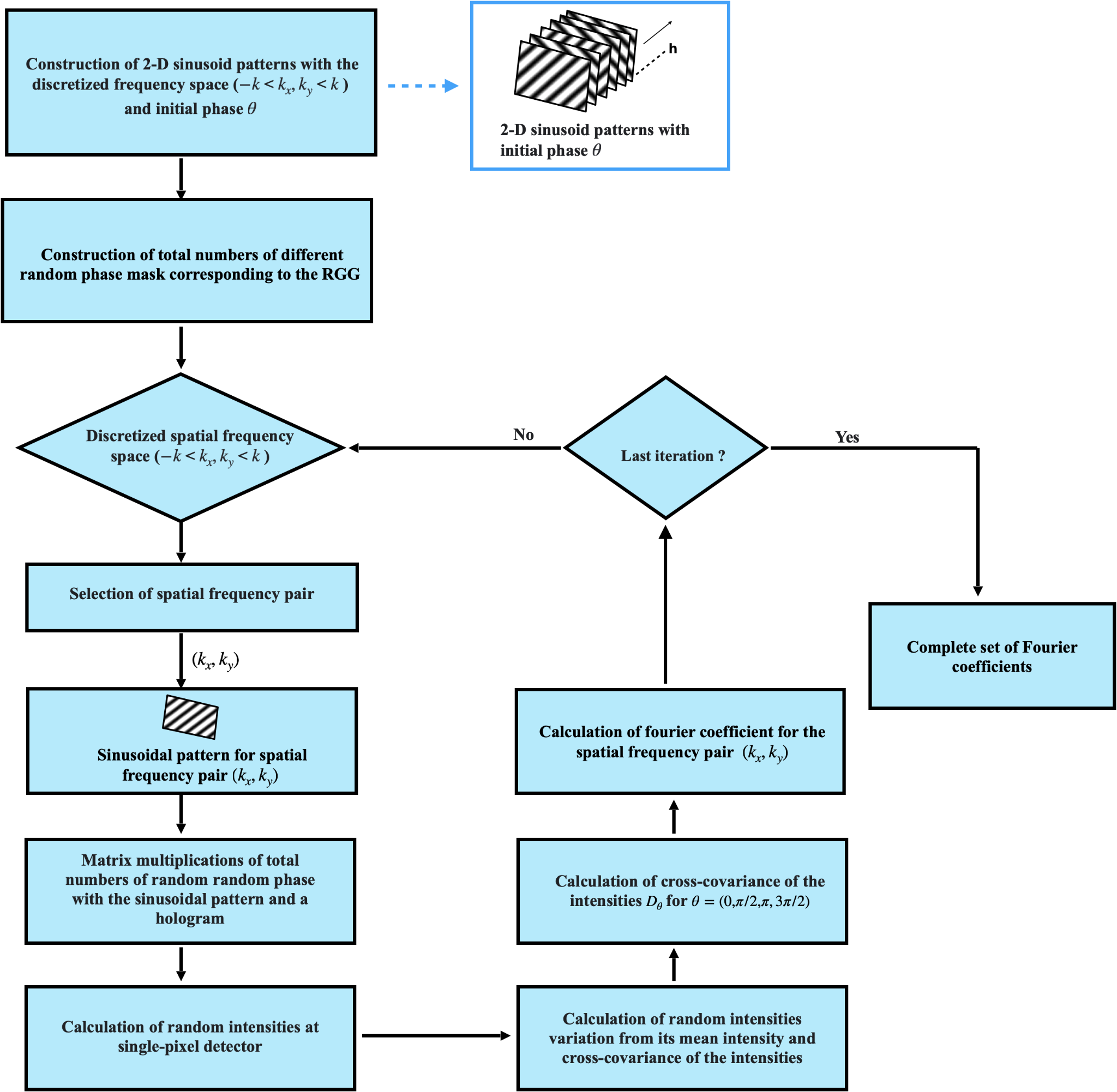}
\caption{Proposed STICH algorithm steps.}
\end{figure}

\section*{Results}
Here, we present computational results processed using MATLAB for the validation of our proposed experimental setup. The phase information of the wavefield is a critical parameter to examine a complex field of an object. Fig 4.(a),(c) and (b),(d) show both amplitude and phase distributions of objects, letter \enquote{P} and number \enquote{3}  respectively, which are directly reconstructed from DH. Applying two-dimensional Fourier transform on the DH brings out three spectra:  a non-modulating central DC term, the desired spectrum, and its off-axis local and unit conjugate. Location of off-axis spectrum is governed by carrier frequency as shown in recording  of DH in Fig. 1(a). The unwanted DC terms containing high-frequency content are digitally suppressed in Fig 4 to highlight the objects located in off-axis position.
Fig 5 and Fig 6 represent the reconstructed complex fields from holograms using the STICH technique at different numbers of random phase masks ($M$) and the quality of reconstruction depends on it. In order to examine the effect of $M$ on reconstruction quality of STICH, we evaluate visibility ($\nu$) and reconstruction
efficiency ($\eta$) \cite{hillman2013digital} for three different numbers of $M = (200, 500, 1000)$ and results are given in Table 1. Amplitude and phase distributions of number \enquote{3} are shown in Fig 5. (a)-(c) and (d)-(f) for $M = (200, 500, 1000)$. Similarly for letter \enquote{P}, Fig 6. (a)-(c) and (d)-(f) show the amplitude and phase distributions for $M = (200, 500, 1000)$. In both object's reconstruction central DC terms are digitally suppressed as shown in Fig 5 and 6. The central DC in the reconstruction appears because use of off-axis hologram as the transparency.
The visibility of a target reconstruction is defined as the degree to which it can be distinguished from background noise. It is calculated as the ratio of the average image intensity level in the signal region to the average background intensity level. Here Otsu's method \cite{otsu1979threshold} is used as a global threshold to identify the signal region.

In order to reconstruct complex fields of size $100\times100$ from hologram of size $200\times200$ using STICH, structured illumination patterns are generated according to Eq. \ref{7}, where $a=0.5$, $b=0.5$, spatial frequencies range is $-2.5\geq(k_{x},k_{y})\leq2.5$ at steps of 0.0505.
\begin{figure}[H]
  \centering
  \begin{tabular}[b]{c}
    \includegraphics[width=.25\linewidth]{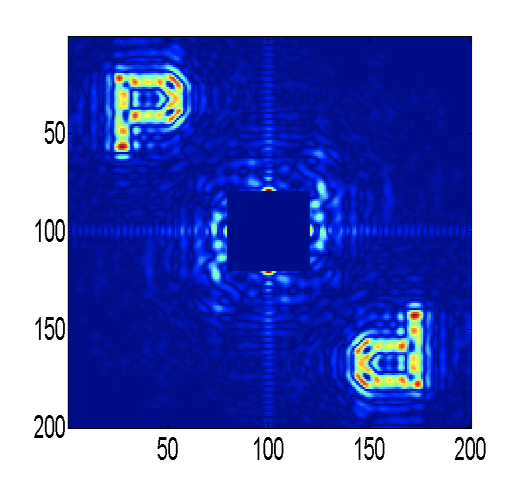} \\
    \small (a)
  \end{tabular} \enspace
  \begin{tabular}[b]{c}
    \includegraphics[width=.25\linewidth]{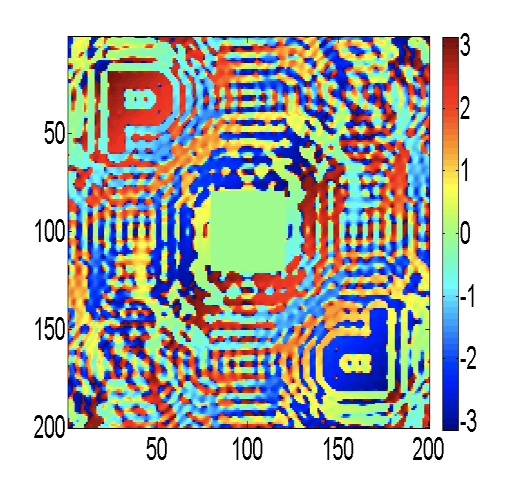} \\
    \small (b)
  \end{tabular} \enspace
  \\
   \begin{tabular}[b]{c}
    \includegraphics[width=.25\linewidth]{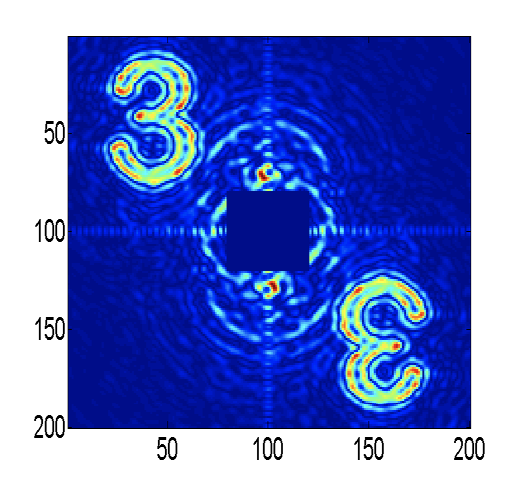} \\
    \small (c)
  \end{tabular}
  \begin{tabular}[b]{c}
    \includegraphics[width=.25\linewidth]{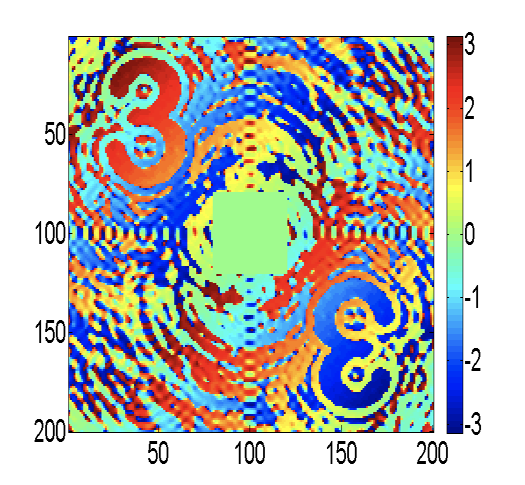} \\
    \small (d)
  \end{tabular}
  \caption{Recovery of complex fields without RGG from holograms of objects, letter \enquote{P} and number \enquote{3} (a) Amplitude distribution of letter \enquote{P} (b) Phase distribution of letter \enquote{P} (c) Amplitude distribution of number \enquote{3} (d) Phase distribution of number \enquote{3}.}
\end{figure}
The calculated visibility value for fig. 4.(a) and (c) are 127.6 and 64.7. The calculated visibility and reconstruction efficiency value for fig. 5.
(a)-(c)  and for fig. 6.(a)-(c) are given in Table 1. (I) and (II) respectively. The upper part of conjugate phase distributions in Fig. 5 and Fig. 6 are highlighted with white color annular rings. From Table 1 and Fig 5 and Fig 6, it can be seen that reconstruction quality improves with increase of value of $M$. 
\begin{figure}[H]
\centering\includegraphics[width=10cm]{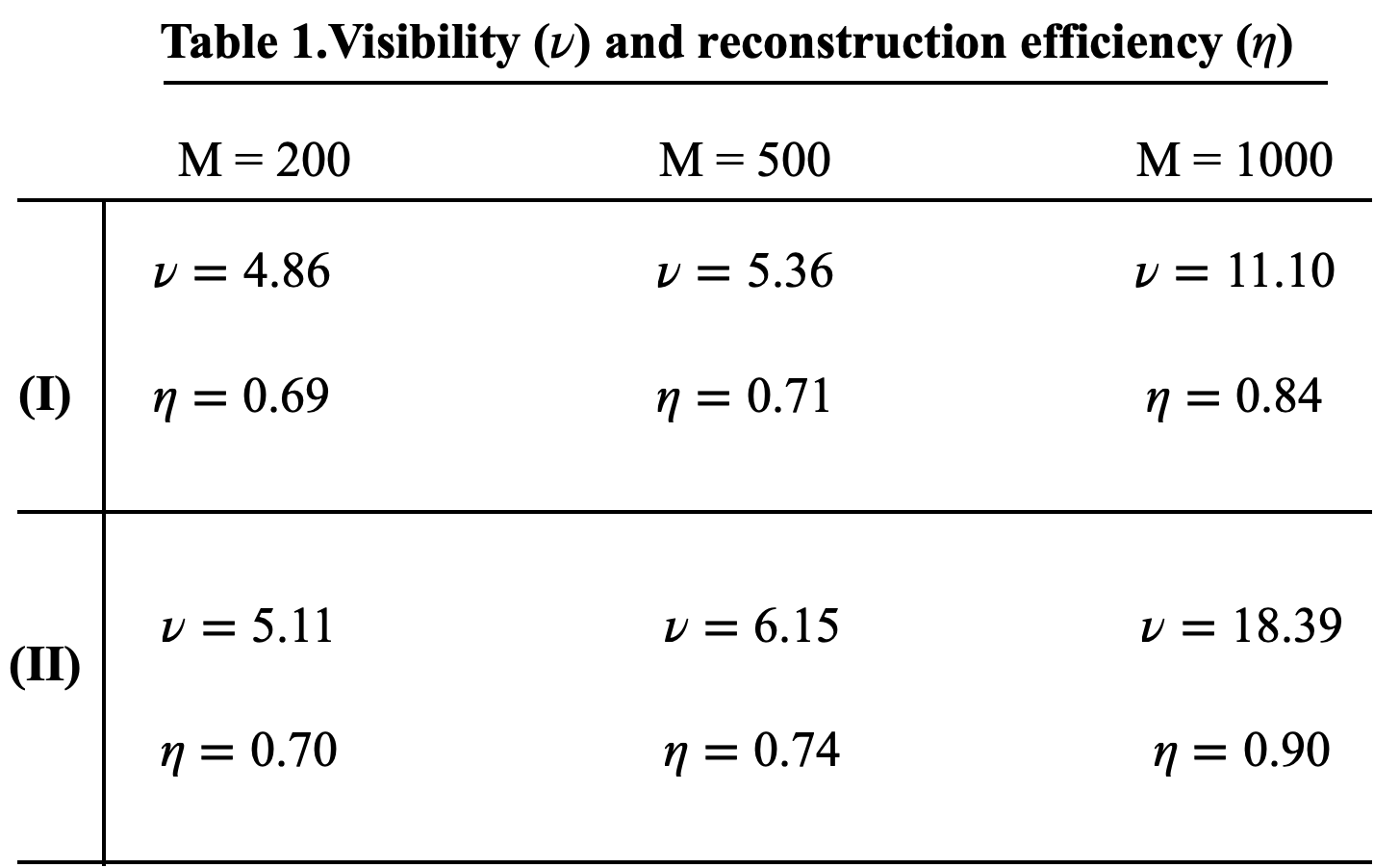}
\end{figure}
\begin{figure}[H]
  \centering
  \begin{tabular}[b]{c}
    \includegraphics[width=.22\linewidth]{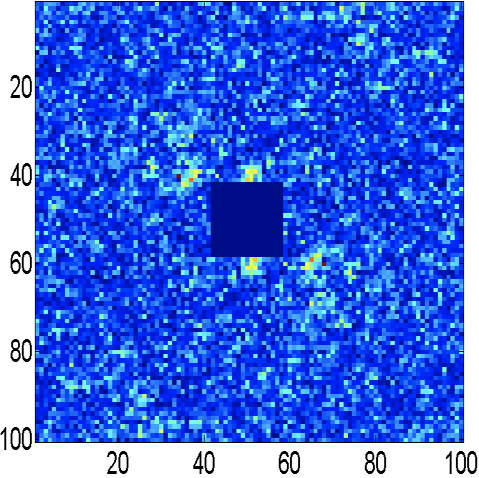} \\
    \small (a)
  \end{tabular} \enspace
  \begin{tabular}[b]{c}
    \includegraphics[width=.22\linewidth]{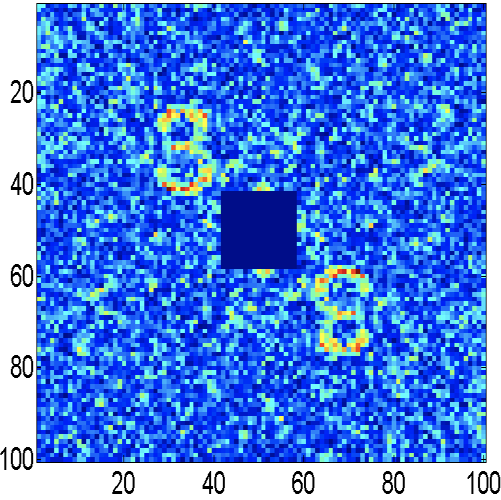} \\
    \small (b)
  \end{tabular} \enspace
   \begin{tabular}[b]{c}
    \includegraphics[width=.22\linewidth]{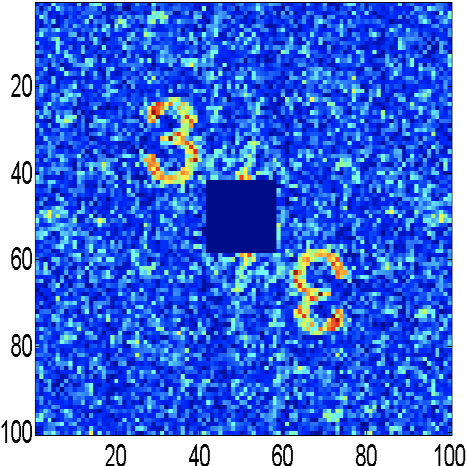} \\
    \small (c)
  \end{tabular}
  \begin{tabular}[b]{c}
    \includegraphics[width=.22\linewidth]{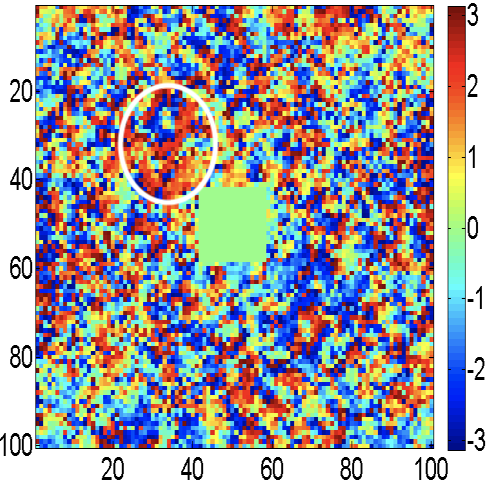} \\
    \small (d)
  \end{tabular} \enspace
  \begin{tabular}[b]{c}
    \includegraphics[width=.22\linewidth]{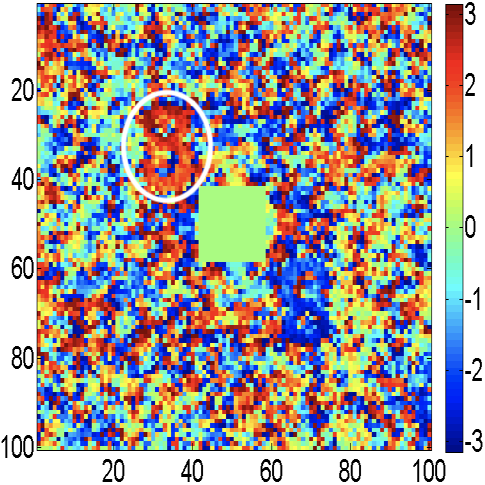} \\
    \small (e)
  \end{tabular} \enspace
   \begin{tabular}[b]{c}
    \includegraphics[width=.22\linewidth]{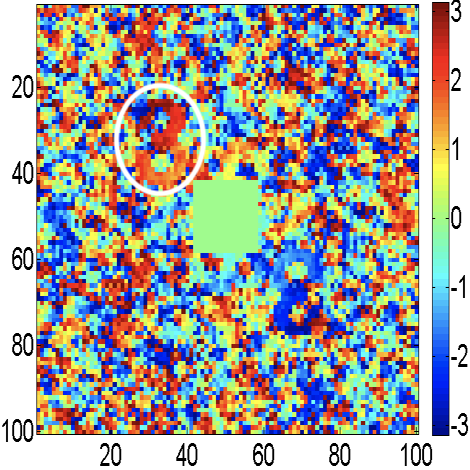} \\
    \small (f)
  \end{tabular}
  \caption{ Recovery of the complex fields using STICH for three different $M$ values: an off-axis hologram of number \enquote{3} is used as transparency (a), (d) Amplitude and phase distribution of object for $M=200$; (b), (e) Amplitude and phase distribution of object for $M=500$; (c), (f) Amplitude and phase distribution of object for $M=1000$.}
\end{figure}
\begin{figure}[H]
  \centering
  \begin{tabular}[b]{c}
    \includegraphics[width=.22\linewidth]{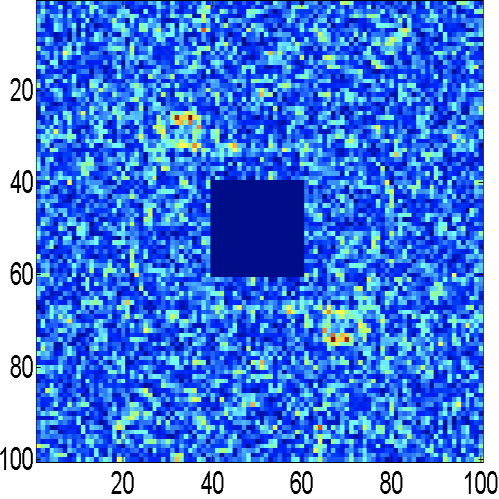} \\
    \small (a)
  \end{tabular} \enspace
  \begin{tabular}[b]{c}
    \includegraphics[width=.22\linewidth]{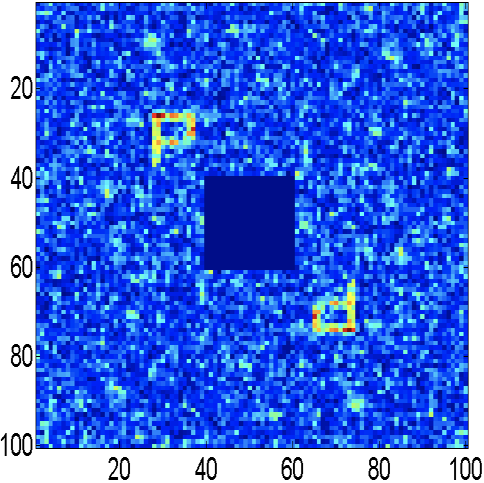} \\
    \small (b)
  \end{tabular} \enspace
   \begin{tabular}[b]{c}
    \includegraphics[width=.22\linewidth]{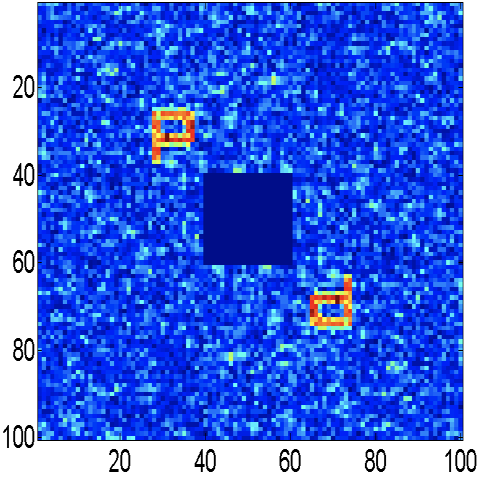} \\
    \small (c)
  \end{tabular}
  \begin{tabular}[b]{c}
    \includegraphics[width=.22\linewidth]{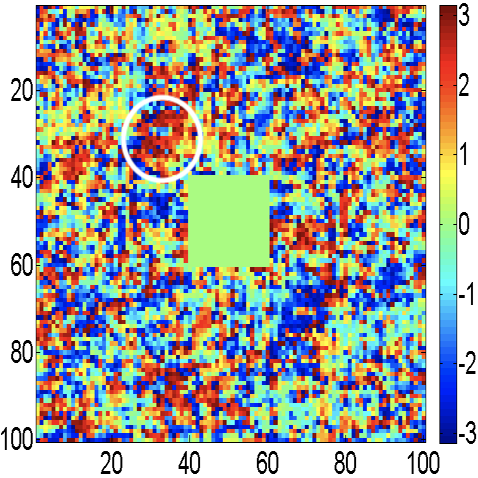} \\
    \small (d)
  \end{tabular} \enspace
  \begin{tabular}[b]{c}
    \includegraphics[width=.22\linewidth]{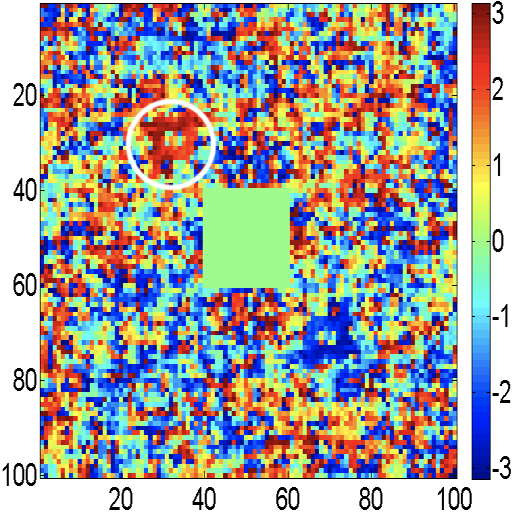} \\
    \small (e)
  \end{tabular} \enspace
   \begin{tabular}[b]{c}
    \includegraphics[width=.22\linewidth]{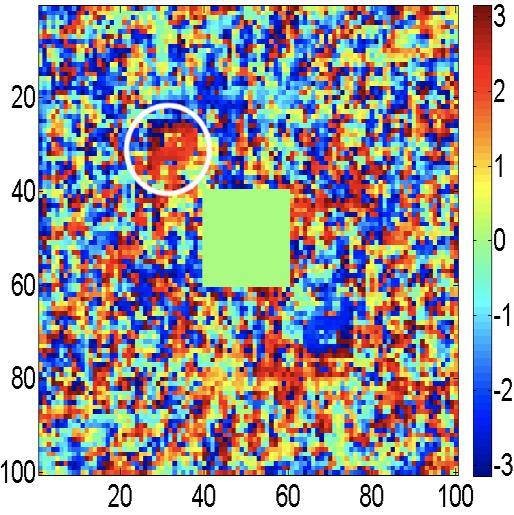} \\
    \small (f)
  \end{tabular}
  \caption{Recovery of the complex fields using STICH for three different $M$ values: an off-axis hologram of letter \enquote{P} is used as transparency (a), (d) Amplitude and phase distribution of object for $M=200$; (b), (e) Amplitude and phase distribution of object for $M=500$; (c), (f) Amplitude and phase distribution of object for $M=1000$.}
\end{figure}
\section*{Conclusion}
In conclusion, a new technique entitled STICH is presented to reconstruct the complex coherence from the intensity measurement with a single-pixel detector and without an interferometric setup. This brings the advantages of compatibility in the reconstruction of complex fields in the correlation-based imaging system. A experimental configuration and computational model of it is described to validate our idea. We have demonstrated the reconstructions of complex fields of objects at different random phase masks and the quality of reconstruction depends on the value of number of random phase masks used to realize the thermal light source. The proposed technique is expected to provide new direction on the coherence holography and imaging through scattering medium.

\bibliography{sample}


\section*{Acknowledgements}
The work is supported by the Science and Engineering Research Board (SERB) India- CORE/2019/000026. T.S acknowledges support from University Grant Commission (UGC), India for his scholarship. 

\section*{Author contributions statement}
A.C.M conceived of idea and build the theoretical basis, experimental design, and completed simulation and preparation of manuscript. T.S involved in  experimental design, simulation, preparation of manuscript. Z.Z provided advice and assistance, reviewing and editing the work. R.K.S involved in supervision, formulation of research goals and aims, funding acquisition, reviewing, and editing.
\end{document}